\documentclass[12pt]{report}
\usepackage{epsfig}
\def\thebibliography#1{\leftline{\large \bf References}\list
  {[\arabic{enumi}]}{\settowidth\labelwidth{[#1]}\leftmargin\labelwidth
    \advance\leftmargin\labelsep
    \usecounter{enumi}}
    \def\newblock{\hskip .11em plus .33em minus .07em}
    \sloppy\clubpenalty4000\widowpenalty4000}

\newcommand{\be}{\begin{eqnarray}}
\newcommand{\ba}{\begin{array}}
\newcommand{\ea}{\end{array}}
\newcommand{\ee}{\end{eqnarray}}
\newcommand{\dslash}{\partial \hskip -0.5em /}

\newcommand{\Tr}{{\rm Tr}}

\newcommand{\A}{{\cal A}}

\newcommand{\imfl}{{\stackrel{\scriptstyle{\rm IMF}}
{\textstyle\longrightarrow}}}

\begin{document}
\rightline{March 1998}
\vskip 2.0truecm
\centerline{\Large\bf NUCLEON STRUCTURE FUNCTIONS FROM A CHIRAL}
\vskip 0.5truecm
\centerline{\Large\bf SOLITON IN THE INFINITE MOMENTUM FRAME$^{*}$}
\baselineskip=15 true pt
\vskip 1.0cm
\centerline{L.\ GAMBERG$^{a)}$, H.\ REINHARDT$^{b)}$
and H.\ WEIGEL$^{b)}$}

\vskip 0.5cm
\centerline{$^{a)}${\it Department of Physics and Astronomy}}
\centerline{\it University of Oklahoma}
\centerline{\it 440 West Brooks Avenue} 
\centerline{\it Norman, Oklahoma 73019--0225, USA}
\vskip 0.5cm
\centerline{$^{b)}${\it Institute for Theoretical Physics}}
\centerline{\it T\"ubingen University}
\centerline{\it Auf der Morgenstelle 14}
\centerline{\it D-72076 T\"ubingen, Germany}
\vskip 1.0cm
\baselineskip=16pt
\centerline{\bf ABSTRACT}
\bigskip
We study the frame dependence of nucleon structure functions 
obtained within a chiral soliton model for the nucleon. Employing light 
cone coordinates and introducing collective coordinates together with 
their conjugate momenta, translational invariance of the solitonic 
quark fields (which describe the nucleon as a localized object) is 
restored. This formulation allows us to perform a Lorentz boost to the 
infinite momentum frame of the nucleon. The major result is that the 
Lorentz contraction associated with this boost causes the leading twist 
contribution to the structure functions to properly vanish when the 
Bjorken variable $x$ exceeds unity. Furthermore we demonstrate that for 
structure functions calculated in the valence quark approximation to 
the Nambu--Jona--Lasinio chiral soliton model the Lorentz contraction 
also has significant effects on the structure functions for moderate 
values of the Bjorken variable $x$.

\vfill
\noindent
$^{*}$
{\footnotesize{Supported in part by the Deutsche Forschungsgemeinschaft
(DFG) under contract Re 856/2-3, the {\it Graduiertenkolleg Hadronen 
und Kerne} at T\"ubingen University, and the U.S. Department of 
Energy (D.O.E.) under contract DE--FE--02--95ER40923.}}
\eject

\normalsize
\baselineskip16pt

\bigskip
\leftline{\large\bf 1. Introduction}
\medskip

Many model calculations of nucleon structure functions are plagued 
by the model being neither translationally invariant nor exhibiting
Lorentz covariance. The reason simply is that the nucleon is described 
as a localized (non--relativistic) object. Such examples are the 
MIT--bag \cite{Ch74}, the center of mass bag \cite{Wa84,So94}
the Skyrme soliton \cite{Sk61,Ad83} and Nambu--Jona--Lasinio 
(NJL\cite{Na61}) chiral soliton \cite{Re88,Al96} models.  
In the context of model structure function calculations this
problem manifests itself as the ``support problem'' 
\cite{So94,Ja75,Hu77,Be87,Be88,Si89,Ja80}. This refers to the 
fact that in these models the structure functions do not vanish 
in the region with Bjorken--$x$ larger than unity.
A number of techniques have been developed to address this problem.  
Sometime ago Jaffe developed a technique that restores Lorentz covariance
in the  $1+1$ bag model while at the same time projecting
both quark and nucleon states onto good momentum \cite{Ja80}.
The merit of this approach is that while solving the support
problem (by simultaneously restoring both translational invariance 
and Lorentz covariance) the corresponding sum rules are 
preserved\footnote{Other (non--relativistic) projection 
techniques, which attempt to generate states with good momentum like 
Peierls--Yoccoz or Peierls--Thouless have been employed in structure 
function calculations \cite{Be87,Be88,Si89,Ar93}. However, recently 
it has been stated \cite{Ja97} that these techniques yield badly 
normalized quark distributions which may be in contradiction with 
established sum rules.}\cite{We96,We97}.

In this paper we extend this technique to a $3+1$ dimension venue
within the NJL chiral soliton model of the nucleon \cite{Re88,Al96}.
In applying this ``projection'' technique to the NJL chiral soliton
one finds the major difference to the bag model is that the 
soliton does not have a {\em boundary} where all densities 
vanish discontinuously; rather the densities decay exponentially in case 
the pion mass is non--zero thus enabling a straightforward 
application of this projection technique.

Generally speaking, denoting by $\Phi(x)$ a localized field 
configuration which, for example solves the classical 
equations of motion in a chiral soliton model,
the above mentioned symmetries can formally be 
restored by defining a projected configuration
\be
\Psi_p(x)= \frac{1}{\cal N}\int d^4 y\ {\rm e}^{-ipy}\ 
S({\bar\Lambda})\Phi\left({\bar\Lambda}^{-1}(x-y)\right)
\label{formproj}
\ee
where ${\bar\Lambda}$ refers to a general Lorentz 
transformation and $S({\bar\Lambda})$ denotes the group operator 
associated with the representation $\Phi$. ${\cal N}$ represents a 
suitable normalization constant. Eq (\ref{formproj}) essentially 
gives the relativistic generalization of the (ordinary) 
non--relativistic projection. The new element is the Lorentz boost 
${\bar \Lambda}$, which restores covariance. Furthermore we emphasize 
that $\Psi_p(x)$ in eq (\ref{formproj}) has the correct transformation 
properties of a field with good four--momentum $p$. Further we note
that if $\Phi(x)$ is a solution to the full time--dependent 
equations of motion in a translationally and Lorentz invariant 
model the configuration 
$S({\bar\Lambda}) \Phi\left({\bar\Lambda}^{-1}(x-y)\right)$ 
will be a solution as well.

In case the localized field configuration is only a solution to 
the static equations of motion ({\it i.e.} time--translational 
invariance is not violated) the integral in eq (\ref{formproj}) 
will only involve the spatial components.

An alternative path to restore the translational symmetry for a static
localized configuration is to use the collective coordinate method of 
Gervais, Jevicki and Sakita \cite{Ge75} introducing a time--dependent 
collective coordinate $\mbox{\boldmath $x$}_0(t)$ which parameterizes 
the spatial position of the localized field 
configuration\footnote{{\it Cf.} ref. \cite{Ra86} for a covariant
approach.}
\be
\Phi_{\mbox{\boldmath $x$}_0}(\mbox{\boldmath $x$})=
\Phi\left(\mbox{\boldmath $x$}-\mbox{\boldmath $x$}_0(t)\right)\ .
\label{push}
\ee
Adopting this field configuration yields a classical Lagrange
function $L(\mbox{\boldmath $x$}_0,
{\dot{\mbox{\boldmath $x$}}_0})$
for the collective coordinate $\mbox{\boldmath $x$}_0(t)$.
This enables one to extract the conjugate momentum 
$\mbox{\boldmath $p$}=\partial L(\mbox{\boldmath $x$}_0,
{\dot{\mbox{\boldmath $x$}}_0})/
\partial{\dot{\mbox{\boldmath $x$}}_0}$,
which is treated as a quantum variable by imposing canonical
commutation relations, {\it i.e.}
$\left[\left(\mbox{\boldmath $x$}_0\right)_i,
\left(\mbox{\boldmath $p$}\right)_j\right]=i\delta_{ij}$.
The wave--function of a nucleon with three--momentum 
$\mbox{\boldmath $p$}$ 
will, for example in the case of a {\em hedgehog} chiral soliton,
be considered a function of the collective coordinates \cite{Br86}
\be
\varphi_{\, \mbox{\footnotesize\boldmath $p$}}
\left(\mbox{\boldmath $x$}_0,t\right)=
\langle \mbox{\boldmath $x$}(t),A(t)|N(p)\rangle=
\frac{1}{2\pi}\sqrt{p_{0}}\
{\rm exp}\left(i\mbox{\boldmath $p$}\cdot 
\mbox{\boldmath $x$}_0\right)\
\sqrt{2I+1}\ {\cal D}^{\hspace{.2cm}I}_{I_{3},-J_{3}}
\left(A(t)\right)
\label{nwfct}
\ee
together with the on--shell condition 
$p_0=\sqrt{\mbox{\boldmath $p$}^2+m^2}$. The ${\cal D}$ function denotes
the projection onto states with good spin ($J,J_3$) and isospin
($I=J,I_3$) which is also required for chiral solitons as discussed 
in section 3. The corresponding collective coordinates 
are comprised in the SU(2) matrix $A(t)$ (see section 3 for a 
brief discussion in the context of the NJL model).
These nucleon states are normalized like Fock states:
$\langle p, I=J | p^\prime, I^\prime=J^\prime\rangle = (2\pi)^3 2p_0\ 
\delta^3\left(\mbox{\boldmath $p$}-\mbox{\boldmath $p$}^\prime\right)
\delta_{I,I^{\prime}}$. 
One may then incorporate Lorentz covariance for nucleon matrix elements 
by appropriately transforming the operator under consideration 
${\cal O}(\mbox{\boldmath $x$}-\mbox{\boldmath $x$}_0,t)$ to a 
suitable frame. This transformation is defined by the boost
$\Lambda_{\mbox{\boldmath $p$}}^{\mu\nu}$
\be
{\cal O}\longrightarrow S\left(\Lambda_{\mbox{\boldmath $p$}}\right)
{\cal O}\Big(\mbox{\boldmath $x$}^\prime-
\mbox{\boldmath $x$}_0^\prime,t^\prime\Big)
S^{-1}\left(\Lambda_{\mbox{\boldmath $p$}}\right)
\quad {\rm where}\quad x^{\prime\mu}=
\left(\Lambda^{-1}_{\mbox{\boldmath $p$}}\right)^\mu_{\ \nu} x^\nu \ .
\label{lboost}
\ee
The important issue now is to identify the relevant frame. When 
{\it e.g.} computing nucleon form factors the Breit (or brick--wall) 
frame is preferred \cite{Ba79}. Here we propose that the 
infinite--momentum--frame (IMF \cite{Ko70}) on the light cone is 
most suited to study deep--inelastic--scattering (DIS). In the 
context of DIS one needs to evaluate nucleon matrix elements of the
form 
\be
f_\Gamma(x)=\lim_{\rm Bj}\  P^{(\Gamma)}_{\mu\nu} 
\int \frac{d^4\xi}{4\pi} \ {\rm exp}\left(iq\cdot\xi \right)
\langle N |\left[J^\mu(\xi),J^\nu(0)\right] |N\rangle \ .
\label{hten}
\ee
Here $\Gamma$ refers to an appropriate spin--flavor matrix,
$P^{(\Gamma)}_{\mu\nu}$ is the associated projector and
$J^\mu(x)$ denotes the hadronic current $J_\mu={\bar \Psi}\Gamma
\gamma_\mu\Psi$. The leading twist contribution to the nucleon
structure functions can be extracted by assuming the Bjorken limit 
\be
-q^2\to\infty \qquad {\rm with} \qquad
x=\frac{-q^2}{2p\cdot q} \quad {\rm fixed}\  ,
\label{bjl1}
\ee
where $q$~denotes the momentum transferred to the nucleon target 
with momentum $p$. In light cone coordinates the IMF is characterized 
by $p^+=(p^0+p^3)/\sqrt{2}\to\infty$.

\bigskip
\leftline{\large\bf 2. Relevance of the Infinite--Momentum--Frame}
\medskip

To begin, we briefly review the definition of light--cone
coordinates with the $x^3$--direction being distinct. Taking an arbitrary
four--vector $x^\mu$ the light--cone coordinates are defined as
\be
x^\pm=\frac{1}{\sqrt{2}}\left(x^0\pm x^3\right)=x_\mp \qquad
\mbox{\boldmath $x$}_\bot = \left(x^1,x^2\right)\ .
\label{lcc1}
\ee
A scalar product of two four--vectors $x$ and $y$ simply reads
\be
x\cdot y = x^- y^+ + x^+ y^- -\mbox{\boldmath $x$}_\bot \cdot
\mbox{\boldmath $y$}_\bot \ .
\label{lcc2}
\ee
When computing structure functions in the Bjorken limit a frame may 
be chosen wherein the spatial components of both the nucleon and 
the photon are along the $x^3$--direction, {\it i.e.}
\be
\mbox{\boldmath $q$}_\bot = 0
\qquad {\rm and} \qquad
\mbox{\boldmath $p$}_\bot = 0\ .
\label{imf1}
\ee
In this frame the Bjorken limit (\ref{bjl1}) becomes rather simple:
\be
q^-\to \infty \qquad {\rm with} \qquad x=-\frac{q^+}{p^+} 
\quad {\rm finite} 
\label{bjl2}
\ee
as long as the reference frame is characterized by $|p^+|>|p^-|$.
A special frame of reference, which satisfies this condition, is the 
IMF defined by
\be
|\mbox{\boldmath $p$}|=p^3\to\infty \ ,
\label{imf2}
\ee
{\it i.e.} the nucleon is moving in the positive $x^3$--direction. 
It is crucial to note that only in the limit 
$|\mbox{\boldmath $p$}|\to\infty$ the parton model interpretation 
of the structure functions is completely consistent. The reason 
being that within the parton model the masses of the partons are 
neglected. This, of course, can only be made consistent with the 
kinematical conditions if the momenta of both the partons and the 
nucleon are large.

In the Bjorken limit the free field anti--commutation relations 
\be
\left\{{\bar \Psi(\xi)},\Psi(0)\right\}=\frac{1}{2\pi}\
\dslash\  \epsilon(\xi_0)\ \delta(\xi^2)
\label{ffcr}
\ee
are adopted to compute the commutator in eq (\ref{hten}). Although
the spinor fields $\Psi$ undergo some (complicated) non--perturbative
interaction this approximation is well justified because in the Bjorken 
limit the intermediate quarks hit by the virtual photon are highly 
off--shell and hence not at all sensitive to the small momenta 
associated with this interaction. The typical momentum scale of this 
interaction is given by the binding of the nucleon. Eq (\ref{ffcr}) 
yields expressions for the structure functions which are 
particularly simple in the light cone formulation (This 
result can be derived in analogy to the calculation presented in 
section 4 below) \cite{Ja85,Si89} 
\be
f_\Gamma(x)=\frac{\sqrt{2}}{4\pi}\ \int d\xi^-
{\rm exp}(-ixp^+\xi^-)
\langle N|\Psi_+^\dagger(\xi^-) \Gamma^2 \Psi_+(0)|
N\rangle_{\xi^+=0, \mbox{\scriptsize\boldmath $\xi$}_\bot=0}  \ ,
\label{lccst1}
\ee
where $\Psi_\pm=\frac{1}{2}\left(\gamma^0\pm\gamma^3\right)\Psi$ 
and $p^+$ is the light--cone component of the nucleon momentum $p$.
Here $\mbox{\boldmath $\xi$}_\bot=0$ results from both the
choice $\mbox{\boldmath $q$}_\bot=0$ for the momentum of the virtual 
photon and in addition from the restriction $\xi^+=0$, which is 
enforced by eq (\ref{ffcr}). The fact that the 
hyperplane $\xi^+=0$ is distinct indicates that when computing 
$f_\Gamma(x)$ in a model which does not exhibit translational 
invariance a frame with $\xi^+=0$ is preferred. We will see in 
section 4 that the boost to the IMF enforces this 
condition. Furthermore it should be noted that in deriving the 
bilinear expression (\ref{lccst1}) a derivative of
$\langle N|\Psi(\xi) \Gamma^2 \Psi(0)|N\rangle$ with respect to
$\xi^+$ has been omitted. As the leading twist contribution of this
matrix element is a polynomial in $\xi\cdot p$ this approximation is
well justified in the IMF since therein $p^-=0$.

Consider now the special case where $\Gamma$ is a projector, 
{\it i.e.} $\Gamma^2=\Gamma$.  When inserting a complete set 
of states $|n\rangle$ and {\it assuming} translational invariance 
the expression (\ref{lccst1}) may be rewritten as
\be
f_\Gamma(x)=
\frac{1}{\sqrt{2}}\sum_n \int \frac{dp_n^3}{4\pi p_n^0}
\delta\left(p^+_n -(1-x)p^+\right) 
\left|\langle n |\Psi^\dagger_+(0)\Gamma|N\rangle \right|^2 \ ,
\label{lccst2}
\ee
where $p_n$ is the eigenvalue of the momentum in state $|n\rangle$
which possess diquark quantum numbers. From the $\delta$--function 
in eq (\ref{lccst2}) one recognizes that because $p_n^+>0$ for 
massive intermediate states the structure functions will vanish
in region $x\approx1-\epsilon$ unless $p_+\to\infty$. Apparently
this just restates that the parton model is only well defined in 
the IMF. Stated otherwise, the numerical integration in (\ref{lccst2}) 
may be difficult because sizable contributions may stem from the region
$p_n^+\to\infty$. 

{}From the preceding discussion on the parton model interpretation of 
the nucleon structure functions we are inclined to conclude that the IMF 
is indeed singled out as the distinct frame to study nucleon structure 
functions in DIS.

\bigskip
\leftline{\large\bf 3. The NJL--Model Chiral Soliton}
\bigskip

Before continuing with the discussion of the IMF we will briefly
review the subject of the chiral soliton in the NJL model. In 
particular the origin of eq (\ref{nwfct}) will be discussed here.
Starting point is the bosonized version of the NJL--model action
\cite{Eb86}
\be
\A&=&\Tr_\Lambda\log(iD)+\frac{1}{4G_{\rm NJL}}
\int d^4x\ {\rm tr}
\left(m^0\left(M+M^{\dag}\right)-MM^{\dag}\right)\ ,
\label{bosact} \\
D&=&i\dslash-\left(M+M^{\dag}\right)
-\gamma_5\left(M+M^{\dag}\right)\ ,
\label{dirac}
\ee
where $M=S+iP$ comprises composite scalar ($S$) and pseudoscalar 
($P$) meson fields. The model parameters (cut--off $\Lambda$, 
coupling constant $G_{\rm NJL}$ and current quark mass matrix $m^0$)
are fixed to reproduce the meson properties, in particular those of 
the pion.

The chiral soliton is given by the hedgehog 
configuration of these meson fields
\be
M_{\rm H}(\mbox{\boldmath $x$})=m\ {\rm exp}
\left(i\mbox{\boldmath $\tau$}\cdot{\hat{\mbox{\boldmath $x$}}}
\Theta(r)\right)\ .
\label{hedgehog}
\ee
In order to compute the functional trace in eq (\ref{bosact}) for this
static configuration a Hamilton operator, $h$ is extracted from the
Dirac operator (\ref{dirac}), {\it i.e.} $D=i\gamma_0(\partial_t-h)$
with
\be
h=\mbox{\boldmath $\alpha$}\cdot\mbox{\boldmath $p$}+m\
{\rm exp}\left(i\gamma_5\mbox{\boldmath $\tau$}
\cdot{\hat{\mbox{\boldmath $x$}}}\Theta(r)\right)\ .
\label{hamil}
\ee
We denote the eigenvalues and eigenfunctions of $h$ by
$\epsilon_\mu$ and $\Psi_\mu$, respectively. In the proper time
regularization scheme the NJL model energy functional is found to
be \cite{Re89,Al96}
\be
E[\Theta]&=&
\frac{N_C}{2}\epsilon_{\rm v}
\left(1+{\rm sign}(\epsilon_{\rm v})\right)
+\frac{N_C}{2}\int^\infty_{1/\Lambda^2}
\frac{ds}{\sqrt{4\pi s^3}}\sum_\nu{\rm exp}
\left(-s\epsilon_\nu^2\right)
\nonumber \\ &&\hspace{6cm}
+m_\pi^2 f_\pi^2\int d^3r  \left(1-{\rm cos}\, \Theta(r)\right) ,
\label{efunct}
\ee
with $N_C=3$ being the number of color degrees of freedom.
The subscript ``${\rm v}$" denotes the valence quark level. This state
is the distinct level bound in the soliton background, {\it i.e.}
$-m<\epsilon_{\rm v}<m$. The chiral angle, $\Theta(r)$, is
obtained by self--consistently extremizing $E[\Theta]$ \cite{Re88}.
States possessing nucleon quantum numbers are generated by 
taking the large amplitude fluctuations (translation and 
rotation) to be time dependent, {\it cf.} eq (\ref{push}),
\be
M(\mbox{\boldmath $x$},t)=
A(t)M_{\rm H}(\mbox{\boldmath $x$}-
\mbox{\boldmath $x$}_0(t))A^{\dag}(t)\ ,
\label{collrot}
\ee
which introduces the collective coordinates discussed in the 
introduction. Upon substitution of the {\it ansatz} (\ref{collrot})
into the action functional (\ref{bosact}) and subsequently expanding to 
quadratic order in the time derivatives of the collective coordinates the
Lagrange function for these coordinates is extracted. It is then 
straightforward to canonically quantize this system. This introduces the 
linear momentum $\mbox{\boldmath $p$}$, the spin $\mbox{\boldmath $J$}$
and isospin $\mbox{\boldmath $I$}$ as the quantities canonical to 
the collective coordinates. In particular the wave--function 
(\ref{nwfct}) is obtained as the eigenfunction of the Hamilton 
operator in the space of the collective coordinates.

Expectation values of quark--bilinears as in eq (\ref{lccst1}) 
are expressed as (regularized) sums over bilinear combinations
of all eigenfunctions $\Psi_\mu$. In practice, however, it turns
out that the dominant contributions to physical quantities stems 
from the distinct valence level $\Psi_{\rm v}$ \cite{Al96}. It 
is therefore reasonable to approximate these bilinears by their
valence quark contribution. In order to obtain nucleon rather than 
soliton structure functions the cranking piece to the wave--function, 
which is induced by the collective rotation $A(t)$, must be included. 
That is, the valence quark wave--function employed to approximate 
the bilinears in the structure functions reads
\be
\Psi_{\rm v}(\mbox{\boldmath $x$},t)
={\rm e}^{-i\epsilon_{\rm v}t}A(t)
\left\{\Psi_{\rm v}(\mbox{\boldmath $x$})
+\frac{1}{2}\sum_{\mu\ne{\rm v}}
\Psi_\mu(\mbox{\boldmath $x$})
\frac{\langle \mu |\mbox{\boldmath $\tau$}\cdot
\mbox{\boldmath $\Omega$}|{\rm v}\rangle}
{\epsilon_{\rm v}-\epsilon_\mu}\right\}\ ,
\label{valrot}
\ee
where the $\Psi_\mu$, which appear within the curly parenthesis, are 
eigenfunctions of the Dirac Hamiltonian
(\ref{hamil}). Upon canonical quantization the angular velocity 
$\mbox{\boldmath $\Omega$}$ is substituted by the nucleon spin 
operator $\mbox{\boldmath $J$}=\alpha^2\mbox{\boldmath $\Omega$}$,
with $\alpha^2$ being the moment of inertia \cite{Re89,Al96}.
The matrix elements of the collective rotations are obtained via
$\langle N | {\rm tr}(\tau_i A \tau_j A^\dagger) |N\rangle=
-(8/3)\langle N | I_i J_j |N\rangle$ \cite{Ad83}. This can 
easily be verified using the wave--function (\ref{nwfct}). In 
what follows we will denote by 
$\varphi_{\, \mbox{\footnotesize\boldmath $p$}}
\left(\mbox{\boldmath $x$}_0,t\right)$ 
only that part of the collective wave--function which depends on the 
tranlational collective coordinate $\mbox{\boldmath $x$}_0$. The 
(iso)rotational degrees of freedom are supposed to be contained 
in the state vector $|N\rangle$.

\bigskip
\leftline{\large\bf 4. Boosting to the Infinite--Momentum--Frame}
\medskip

Here we present the major topic of this paper, namely the calculation
of the chiral soliton model structure function in the IMF. This 
calculation is motivated by Jaffe's conjecture \cite{Ja80}; namely
that the expressions for structure functions with proper support,
obtained in a covariant $1+1$--dimensional model, can formally be 
transferred to a realistic $3+1$--dimensional model, which describes
the nucleon as an extended object. When addressing this conjecture 
within the context of a chiral soliton model it is crucial to note 
that, as the soliton is a continuous field configuration, problems 
originating from the existence of the bag boundary ({\it e.g.} 
periodic boundary conditions and a fluctuating bag boundary 
\cite{Ja80,Be87}) do not occur.

In light--cone coordinates the boost from the rest frame to the IMF
in $x^3$--direction may be parameterized in terms of the rapidity 
$\Omega$:
\be
p^+={\rm e}^{+\Omega}\left(\frac{m}{\sqrt{2}}\right)\ , \quad
p^-={\rm e}^{-\Omega}\left(\frac{m}{\sqrt{2}}\right)\ , \quad
\mbox{\boldmath $p$}_\bot = 0 \ ,
\label{rapidity}
\ee
where $m$ refers to the invariant mass of the nucleon. Apparently 
the IMF is characterized by the limit $\Omega\to\infty$. In this 
limit the transformation matrix for Dirac spinors becomes
\be
S(\Lambda)&=&\sqrt{\frac{p^0+m}{2m}}\mathbf{1} + 
\sqrt{\frac{p^0-m}{2m}}\alpha_3 
\nonumber \\
&\imfl& S(\Lambda_\Omega) +{\cal O}\left(\frac{m}{p^+}\right)
\quad {\rm with} \quad
S(\Lambda_\Omega)= \frac{1}{2}\ 
{\rm exp}\left(\frac{\Omega}{2}\right)
\left(\mathbf{1}+\alpha_3\right)\ ,
\label{dirtrans}
\ee
where $\alpha_3$ denotes a Dirac matrix. In what follows we will
consider localized quark spinors, which result from some static soliton 
calculation \cite{Al96}, boosted to the IMF
\be
\Psi(\xi)&\longrightarrow& S(\Lambda)
\Psi\left(\Lambda^{-1}(\xi-x_0)\right)
\nonumber \\
&\imfl& \ \frac{1}{2} \left(\mathbf{1}+\alpha_3\right)
\Psi\left({\rm e}^\Omega(\xi^- -x^-_0),
\mbox{\boldmath $\xi$}_\bot-\mbox{\boldmath $x$}_{0\bot},
(\xi^+ - x^+_0)=0\right)
{\rm exp}\left(\frac{\Omega}{2}\right)\ .
\label{dirimf}
\ee
Note that the $\xi^-$ coordinate acquires the factor ${\rm exp}(\Omega)$
due to the appearance of $\Lambda^{-1}$ in the argument. Here the 
collective coordinate $x_0^\mu=(0,\mbox{\boldmath $x$}_0)$, which 
labels the position of the soliton has been introduced, {\it cf} eq 
(\ref{push}). As noted before, this collective coordinate serves to 
generate states of good momentum $p$ from a configuration which is 
not translationally invariant. Eventually an integration over this 
coordinate is performed. Furthermore we have written the rapidity to 
the right of the wave--function. Later this ordering will be important 
when elevating $p^+$ as an operator in the space of the collective 
coordinates.

In the discussion above we emphasized that the frame with $\xi^+=0$, 
{\it i.e.} the null plane, was distinct for structure function 
calculations. It is hence the IMF ($p^+\to\infty)$ which is
preferred, {\it i.e.} to leading order in $m/p^+$,
this condition is satisfied because 
$\xi^{\prime+}={\rm e}^{-\Omega}\xi^+\to0$. This has already been 
used in eq (\ref{dirimf}).

Assuming again the free field commutation relations (\ref{ffcr}) for 
the Dirac fields in the defining equation for the hadronic tensor
(\ref{hten}) we generalize the rest frame expression of the 
forward moving intermediate quark to the structure function $F_{1}^{(+)}(x)$ 
\cite{Ja75,Hu77} by substituting the boosted quark wave function 
(\ref{dirimf}) (see appendix A for more details)\footnote{Here the 
projector simply is $P_{\mu\nu}=g_{\mu\nu}$.}, 
\be
F^{(+)}_1\left(x,q^2\right)
\hspace{-0.1cm}&=&\hspace{-0.1cm}-\hspace{-0.1cm}
\lim_{q^-\to\infty}\frac{q^\rho}{4\pi}
\int d\xi^+d\xi^-d^2\xi_\bot\ d^3x_0 \
{\rm exp}\left[i\left(q^+\xi^-+q^-\xi^+\right)\right]
\frac{\delta(\xi^2)\epsilon(\xi^+ + \xi^-)}{2\pi}
\nonumber \\ && \hspace{1.0cm}\times
\Big\langle p^+, p^-=\frac{m}{\sqrt{2}p^+},
\mbox{\boldmath $p$}_\bot=0 \Big|\ 
\mbox{\boldmath $x$}_{0}\Big\rangle
\nonumber \\ && \hspace{1.0cm}\times
\langle N|\Bigg[S(\Lambda_\Omega)\Psi\left({\rm e}^\Omega(\xi^- -x^-_0),
\mbox{\boldmath $\xi$}_\bot-\mbox{\boldmath $x$}_{0\bot},
(\xi^+ - x^+_0)=0\right)\Bigg]^\dagger \Gamma^\rho
\nonumber \\ && \hspace{4.0cm}\times
S(\Lambda_\Omega)\Psi\left({\rm e}^\Omega(-x^-_0),
-\mbox{\boldmath $x$}_{0\bot},x_0^+=0\right)\Big|N\rangle
\nonumber \\ && \hspace{1.0cm}\times
\langle \mbox{\boldmath $x$}_{0}\ 
\Big| p^+, p^-=\frac{m}{\sqrt{2}p^+},
\mbox{\boldmath $p$}_\bot=0\Big\rangle\ .
\label{f1p1}
\ee
The contribution of the backward moving quark,
$F^{(-)}_1(x,q^2)$, comes with the opposite sign and the arguments 
of the spinors exchanged. For the ongoing discussion it is, however,
sufficient to only consider $F^{(+)}_1(x,q^2)$. Here we treat 
$x_0=\hat{x}_0$ as an operator in the space of the collective 
coordinates in the sense that 
$f(\hat{x}_0)|x_0\rangle=f(x_0)|x_0\rangle$. Of course, we should take
$x_0^-=-x_0^3/\sqrt{2}$ as there is no collective coordinate in
the time component since the soliton preserves time--translational
invariance (when employing light cone coordinates we omit the 
vector notation).

Now it is appropriate to express the arguments of the spinors in 
terms of integrals over $\delta$--functions and Fourier--expand 
the latter. This introduces dummy variables $\zeta_\mu$ and 
$\zeta_\mu^\prime$ as well as their conjugates $\alpha$ 
and $\alpha^\prime$
\be
F^{(+)}_1\left(x,q^2\right)&=&
-\lim_{q^-\to\infty}\
\frac{q^\rho}{8\pi^2}\frac{m^2}{2}\ {\rm exp}\left(\Omega\right)
\nonumber \\ && \hspace{-1cm}\times
\int \frac{dx_0^-}{\sqrt2}d^2x_{0\bot}
\Big(\int d\xi^+d\xi^-d^2\xi_\bot\Big)
{\rm exp}\left[i\left(q^+\xi^-+q^-\xi^+\right)\right]
{\delta(\xi^2)\epsilon(\xi^+ + \xi^-)}
\nonumber \\ && \hspace{-1cm} \times
\int d\zeta^- d\zeta^{\prime-}
d^2\zeta_\bot d^2\zeta_\bot^\prime\
\frac{d\alpha}{2\pi}\frac{d\alpha^\prime}{2\pi}\
\delta^2\left(\mbox{\boldmath $\xi$}_\bot-
\mbox{\boldmath $x$}_{0\bot}-
\mbox{\boldmath $\zeta$}^\prime_\bot\right)
\delta^2\left(-\mbox{\boldmath $x$}_{0\bot}-
\mbox{\boldmath $\zeta$}_\bot\right)
\langle p | x_0\rangle
\nonumber \\ && \hspace{-1cm}\times
\langle N | \Psi^\dagger_+(\zeta^\prime)\
{\rm exp}\left[i\alpha^\prime\left(
p^+\xi^-+\Lambda-\frac{m}{\sqrt2}\zeta^{\prime-}\right)\right]
\Gamma_\rho
\nonumber \\&& \hspace{1cm}\times
\Psi_+(\zeta)\ {\rm exp}
\left[-i\alpha\left(\Lambda-\frac{m}{\sqrt2}\zeta\right)\right]
|N\rangle \langle x_0|p\rangle \ .
\label{f1p2}
\ee
Here we have made explicit the integration over the collective 
coordinate $x_0$ as well as the Lorentz boost $S(\Lambda_\Omega)$. 
Also we have split the nucleon state in a piece containing the 
collective momentum $|p\rangle$ and the remaining degrees 
of freedom $|N\rangle$ like {\it e.g.} isospin, see also 
eq (\ref{nwfct}). In addition we have introduced the ordering
\be
\Lambda = -x_0^-p^+ = 
-\frac{1}{2}\left(x_0^-p^+ + p^+x_0^-\right) \ .
\label{lcboost}
\ee
When implementing Lorentz covariance in the $x^3$--direction, 
which is sufficient since we consider nucleon states with 
$\mbox{\boldmath $p$}_\bot=0$, this object as well as 
the momentum $p^+$ have to be considered as operators in the 
space of the collective coordinates. For this reason we have 
not only taken the Hermitian ordering but also carefully 
treated the ordering of the collective wave--functions 
$\langle x_0 | p\rangle$. As $p^+$ and $x_0^-$ are conjugate 
to each other we have the commutation relations
\be
[x_0^-,p^+]=-i \quad {\rm and} \quad
[\Lambda,p^+]=ip^+  \ .
\label{comrel}
\ee
Imposing these relations should be considered as the 
semiclassical quantization of the translational degrees of freedom 
for the classical soliton configuration. It is the analogue of the 
cranking approach to generate states of good spin and isospin 
\cite{Ad83} for the chiral soliton, which is nothing but the Lagrange
form of the collective coordinate method \cite{Ge75}. Care has to 
be taken when computing the structure 
function (\ref{f1p2}) because these operators act on the collective
wave--function $\langle x_0|p\rangle$. At this point also the 
ordering chosen in (\ref{dirimf}) is crucial. Having put
${\rm exp}(\Omega/2)=p^+/2\sqrt{2}m$ to the right of the quark 
wave--function not only ensures that the boost indeed goes with 
the nucleon momentum in the IMF but also provides consistent 
normalization of quark and nucleon wave--functions\footnote{We 
could as well have chosen an ordering wherein ${\rm exp}(\Omega/2)$
would have been to the left of the quark wave--function in eq 
(\ref{dirimf}). In that case a consistent normalization would 
require to introduce a {\it scale dependent} mass 
${\rm exp}(\Omega/2)=p^+/2\sqrt{2}\ {\rm exp}(\alpha^{(\prime)})m$. 
The final result for the structure functions would remain 
unchanged. This consideration, however, also indicates that the 
restoration of Lorentz covariance can only be accomplished in the 
subspace of the nucleon ground state. A simultaneous treatment of 
excited baryons does not seem to be feasible at present.}. We already 
made use of this ordering 
when simply writing ${\rm exp}(\Omega)$ in eq (\ref{f1p2}). 

On repeated application of the commutation relations (\ref{comrel}) 
one finds the operator identities
\be
{\rm exp}\left[i\alpha\left(p^+\xi^-+\Lambda\right)\right]&=&
{\rm exp}\left(ip^+\xi^-\right)
{\rm exp}\left(i\alpha\Lambda\right)
{\rm exp}\left(-ip^+\xi^-\right)
\label{transl}\\
{\rm exp}\left(-ip^+\xi^-\ {\rm e}^{-\alpha}\right)&=&
{\rm exp}\left(i\alpha\Lambda\right)
{\rm exp}\left(-ip^+\xi^-\right)
{\rm exp}\left(-i\alpha\Lambda\right) \ .
\label{boost}
\ee
It is very instructive to discuss the physical content of these
equations. First we note that the unitary operator 
${\rm exp}\left(ip^+\xi^-\right)$ generates translations on the 
light cone, {\it i.e.} $x_0^-$, by the amount of $\xi^-$. As the boost 
operator $\Lambda$ is linear in $x_0^-$ any function $f(\Lambda)$ 
transforms as 
$${\rm exp}\left(ip^+\xi^-\right)f(\Lambda)\
{\rm exp}\left(-ip^+\xi^-\right)=f(\Lambda+p^+\xi^-) \ . $$ 
Eq (\ref{transl}) corresponds to 
$f(\Lambda)={\rm exp}\left(i\alpha\Lambda\right)$. The relation 
(\ref{transl}) furthermore ensures that a shift ($\delta\xi_\mu$) 
in the coordinate $\xi_\mu$ just adds the phase 
${\rm exp}(ip^+\delta\xi^-)$. Of course, this shows that we 
have restored translational invariance in the subspace under 
consideration which is characterized by 
$\mbox{\boldmath $p$}_\bot=0$.
Similarly the operator ${\rm exp}\left(-i\alpha\Lambda\right)$
generates boosts with the rapidity $\alpha$:
$${\rm exp}\left(-i\alpha\Lambda\right)g(p^+)\
{\rm exp}\left(i\alpha\Lambda\right)=
g\left({\rm e}^\alpha p^+\right)\ . $$
Eq (\ref{boost}) is obtained for 
$g(p^+)={\rm exp}\left(-ip^+\xi^-\right)$. Hence by imposing 
the commutation relations (\ref{comrel}) in the space 
of the collective coordinates we have implemented the correct 
transformation properties of the localized Dirac spinor for the 
problem at hand.

We are now in a position to calculate the integrals over the 
dummy variables $d\zeta^-$,...., $d\alpha^\prime$. We find
\be
&&\int d\zeta^- d\zeta^{\prime-}
d^2\zeta_\bot d^2\zeta_\bot^\prime\
\frac{d\alpha}{2\pi}\frac{d\alpha^\prime}{2\pi}\
\delta^2\left(\mbox{\boldmath $\xi$}_\bot-
\mbox{\boldmath $x$}_{0\bot}-
\mbox{\boldmath $\zeta$}^\prime_\bot\right)
\delta^2\left(-\mbox{\boldmath $x$}_{0\bot}-
\mbox{\boldmath $\zeta$}_\bot\right)
\nonumber \\ &&\hspace{1cm}\times
{\rm exp}\left[ip^+\xi^-(1-e^{-\alpha^\prime})\right]
\varphi^\dagger_{p^+}\left(x_0^-,\mbox{\boldmath $x$}_{0\bot}\right)
{\rm exp}\left[i(\alpha^\prime-\alpha)\Lambda\right]
\varphi_{p^+}\left(x_0^-,\mbox{\boldmath $x$}_{0\bot}\right)
\nonumber \\ &&\hspace{1cm}\times
\langle N |\Psi^\dagger_+\left(\zeta^\prime\right)\Gamma_\rho
\Psi_+\left(\zeta\right)|N\rangle
{\rm exp}\left[i\left(\epsilon(\zeta_0^\prime-\zeta_0)
-\frac{m}{\sqrt{2}}\left(\alpha^\prime\zeta^{\prime-}-
\alpha\zeta\right)\right)\right]
\label{f1p3}
\ee
Here $\epsilon$ denotes the energy eigenvalue of $\Psi$ which
is determined from the static Dirac equation. Once again we make 
use of $\Lambda$ being the boost operator in the space of the 
collective coordinate to perform the integral over the collective 
coordinate $x_0$
\be
&&\int \frac{dx_0^-}{\sqrt{2}}d^2x_{0\bot}\
\varphi^\dagger_{p^+}\left(x_0^-,\mbox{\boldmath $x$}_{0\bot}\right)
{\rm exp}\left[i(\alpha^\prime-\alpha)\Lambda\right]
\varphi_{p^+}\left(x_0^-,\mbox{\boldmath $x$}_{0\bot}\right)
\delta^2\left(\mbox{\boldmath $\xi$}_\bot-
\mbox{\boldmath $x$}_{0\bot}-
\mbox{\boldmath $\zeta$}^\prime_\bot\right)
\nonumber \\ &&\hspace{2cm}
=\int \frac{dx_0^-}{\sqrt{2}}
\varphi^\dagger_{p^+}\left(x_0^-,\mbox{\boldmath $x$}_{0\bot}\right)
\varphi_{p^{\prime +}}\left(x_0^-,\mbox{\boldmath $x$}_{0\bot}\right)
\Big|_{\mbox{\scriptsize\boldmath $x$}_{0\bot}=
\mbox{\scriptsize\boldmath $\xi$}_\bot
-\mbox{\scriptsize\boldmath $\zeta$}_\bot
\ ,\ p^{\prime +}={\rm e}^{(\alpha^\prime-\alpha)}p^+}
\nonumber \\ &&\hspace{2cm}
=2\pi\ 2p^+ \delta\left(p^+-{\rm e}^{(\alpha^\prime-\alpha)}p^+\right)
\Big|_{\mbox{\scriptsize\boldmath $x$}_{0\bot}=
\mbox{\scriptsize\boldmath $\xi$}_\bot
-\mbox{\scriptsize\boldmath $\zeta$}_\bot} \ .
\label{collint}
\ee
This $\delta$--function enforces $\alpha=\alpha^\prime$ which
removes the (c--number) ambiguity stemming from the various 
definitions of the boost operator (\ref{lcboost}).
Substituting the above results into eq (\ref{f1p2}) yields
\be
F^{(+)}_1\left(x,q^2\right)\hspace{-0.2cm}&=&\hspace{-0.1cm}
-\lim_{q^-\to\infty}\
\frac{q^\rho m p^+}{4\sqrt{2}\pi^2}\int d\xi^+d\xi^-d^2\xi_\bot
{\rm exp}\left[i\left(q^+\xi^-+q^-\xi^+\right)\right] 
\delta(\xi^2)\ \epsilon(\xi^+ + \xi^-)
\nonumber \\ && \hspace{-1cm} \times
\int dz^- d^2z_\bot d\gamma^- d^2\gamma_\bot \frac{d\alpha}{2\pi}
\delta^2\left(\mbox{\boldmath $\gamma$}_\bot-
\mbox{\boldmath $x$}_\bot\right)
{\rm exp}\left[\frac{i}{\sqrt{2}}\left(\epsilon-m\alpha\right)
\gamma^-\right]
\nonumber \\ && \hspace{-1cm} \times
{\rm exp}\left[ip^+x^-\left(1-e^{-\alpha}\right)\right]
\langle N |\Psi^\dagger_+\left(z+\frac{\gamma}{2}\right)\Gamma_\rho
\Psi_+\left(z-\frac{\gamma}{2}\right)|N\rangle\ ,
\label{f1p4}
\ee
where we have changed the dummy variables to $\gamma=\zeta^\prime-\zeta$
and $z=(\zeta+\zeta^\prime)/2$. The expression (\ref{f1p4}) can further 
be simplified by introducing the Fourier transform of the Dirac spinor
\be
\Psi\left(\mbox{\boldmath $y$}_\bot,
y_3=- \frac{y^-}{\scriptstyle\sqrt{2}}\right)=
\int \frac{d^2k_\bot dk_3}{2\pi^2}\
{\rm exp}\left[i\left(\frac{k_3y^-}{\scriptstyle\sqrt{2}}-
\mbox{\boldmath $k$}_\bot\cdot
\mbox{\boldmath $y$}_\bot\right)\right]
{\tilde \Psi}\left(\mbox{\boldmath $k$}_\bot,k_3\right) \ .
\label{ftrans}
\ee
This now allows us to integrate over the dummy variable $z$
\be
F^{(+)}_1\left(x,q^2\right)&=&
-\lim_{q^-\to\infty}\
\frac{q^\rho m p^+}{\pi}\int d\xi^+d\xi^-d^2\xi_\bot
{\rm exp}\left[i\left(q^+\xi^-+q^-\xi^+\right)\right]
\delta(\xi^2)\ \epsilon(\xi^+ + \xi^-)
\nonumber \\ && \hspace{1cm} \times
\int d\gamma^- d^2\gamma_\bot \frac{d\alpha}{2\pi}\
{\rm exp}\left[ip^+x^-\left(1-e^{-\alpha}\right)\right]
\delta^2\left(\mbox{\boldmath $\gamma$}_\bot-
\mbox{\boldmath $x$}_\bot\right)
\label{f1p5} \\ && \hspace{-2cm} \times
\int \frac{d^2k_\bot dk_3}{2\pi^2}\
{\rm exp}\left[\frac{i}{\sqrt{2}}
\left(\epsilon-m\alpha-k_3\right)\gamma^- +
i\mbox{\boldmath $\gamma$}_\bot\cdot\mbox{\boldmath $k$}_\bot\right]
\langle N |{\tilde \Psi}^\dagger_+
\left(\mbox{\boldmath $k$}\right)
\Gamma_\rho{\tilde \Psi}_+
\left(\mbox{\boldmath $k$}\right)|N\rangle\ .
\nonumber
\ee
The additional factor $1-{\rm e}^{-\alpha}$ in the exponential is 
nothing but the Lorentz contraction associated with the boost into 
the IMF. Its appearance is crucial and may be interpreted as an effect 
associated with the relativistic recoil. 

We may now integrate over the coordinate $\xi$ by treating 
$\delta(\xi^2)$ in the well--known manner ({\it cf.} ref. 
\cite{Ja85}). This enforces $\xi^+=0$ as well as 
$\mbox{\boldmath $\xi$}_\bot=0$. The latter than allows us to also 
perform the $\mbox{\boldmath $\gamma$}_\bot$ integral
\be
F^{(+)}_1\left(x,q^2\right)\hspace{-0.2cm}&=&
-mp^+\int \frac{d\alpha}{2\pi} d\xi^-
{\rm exp}\left[-ip^+\xi^-\left(x-1+{\rm e}^{-\alpha}\right)\right]
\label{f1p6} \\ && \hspace{-1.0cm} \times
\int d\gamma^- \frac{d^2k_\bot dk_3}{2\pi^2}\
{\rm exp}\left[\frac{i}{\sqrt{2}}
\left(\epsilon-m\alpha-k_3\right)\gamma^-\right]
\langle N |{\tilde \Psi}^\dagger_+
\left(\mbox{\boldmath $k$}\right)
\Gamma^+{\tilde \Psi}_+
\left(\mbox{\boldmath $k$}\right)|N\rangle\ .
\nonumber
\ee
Here we have also introduced the Bjorken variable via $q^+=-xp^+$ in the 
IMF, see eq (\ref{bjl2}). At this point we recognize that the integration 
over $\xi^-$ will yield $\delta(x-1+{\rm e}^{-\alpha})$ which guarantees 
that $F^{(+)}_1(x,q^2)$ vanishes for $x>1$.  It is important to note
that this $\delta$--function will also appear in case the IMF condition 
$p^+\to\infty$ is relaxed. However, the remainder of the matrix 
element will be more complicated and the $\alpha$--integration, which 
we perform in the next step, might become infeasible. Performing 
now the $\alpha$--integration yields
\be
F^{(+)}_1\left(x,q^2\right)&=&\frac{m}{1-x}
\int d\gamma^- \frac{d^2k_\bot dk_3}{2\pi^2}\
{\rm exp}\left[\frac{i}{\sqrt{2}}\Big(m\ {\rm ln}(1-x)
+\epsilon-k_3\Big)\gamma^-\right]
\nonumber \\ && \hspace{2.0cm} \times
\langle N |{\tilde \Psi}^\dagger_+
\left(\mbox{\boldmath $k$}\right)
\Gamma^+{\tilde \Psi}_+
\left(\mbox{\boldmath $k$}\right)|N\rangle\ .
\label{f1p7}
\ee
The $\gamma^-$ integral yields yet another $\delta$--function which 
fixes $k_3=m{\rm ln}(1-x)+\epsilon$. Introducing spherical coordinates 
in momentum space and making the Dirac matrix $\Gamma^+$ explicit we 
arrive at
\be
\hspace{-0.3cm}
F^{(+)}_1\left(x,q^2\right)&\hspace{-0.1cm}=\hspace{-0.1cm}&
\frac{m}{\pi(1-x)}
\int_{k_{\rm min}}^\infty \hspace{-0.2cm} kdk d\varphi 
\langle N |{\tilde \Psi}^\dagger
\left(\mbox{\boldmath $k$}\right)
\left(1+\alpha_3\right){\tilde \Psi}
\left(\mbox{\boldmath $k$}\right)|N\rangle 
\Big|_{{\rm cos}\Theta=
{\textstyle\frac{m\ {\rm ln}(1-x)+\epsilon}{k}}}
\label{f1p8}
\ee
with $k_{\rm min}=|m\ {\rm ln}(1-x)+\epsilon|$.
Application of the parity transformation finally leads to a 
very familiar expression \cite{Ja80}
\be
\hspace{-0.3cm}
F^{(+)}_1\left(x,q^2\right)&\hspace{-0.1cm}=\hspace{-0.1cm}&
\frac{m}{\pi(1-x)}
\int_{k_{\rm min}}^\infty \hspace{-0.2cm} kdk d\varphi \
\langle N |{\tilde \Psi}^\dagger
\left(\mbox{\boldmath $k$}\right)
\left(1-\alpha_3\right){\tilde \Psi}
\left(\mbox{\boldmath $k$}\right)|N\rangle
\Big|_{{\rm cos}\Theta=-
{\textstyle \frac{m\ {\rm ln}(1-x)+\epsilon}{k}}} \ .
\label{f1p9}
\ee
Apparently the leading order in a large $p^+$ expansion does not 
depend on $p^+$ itself. Without the boost into the IMF we had found 
previously \cite{We96}
\be
\hspace{-0.3cm}
F^{(+)}_1\left(x,q^2\right)&\hspace{-0.1cm}=\hspace{-0.1cm}&
\frac{m}{\pi}\int_{k_{\rm min}}^\infty\hspace{-0.2cm}kdk d\varphi \
\langle N |{\tilde \Psi}^\dagger
\left(\mbox{\boldmath $k$}\right)
\left(1-\alpha_3\right){\tilde \Psi}
\left(\mbox{\boldmath $k$}\right)|N\rangle
\Big|_{{\rm cos}\Theta=
{\textstyle \frac{mx-\epsilon}{k}}} 
\label{f1p10}
\ee
together with the lower boundary $k_{\rm min}=|mx-\epsilon|$.
Comparing eqs (\ref{f1p9}) and (\ref{f1p10}) we finally establish 
that the structure functions which are calculated from a static 
localized quark configuration transform as
\be
f_{\rm IMF}(x)=\frac{\Theta(1-x)}{1-x} f_{\rm RF}
\Big(-{\rm ln}(1-x)\Big)
\label{fboost}
\ee
when the quark fields are boosted from the nucleon rest frame (RF)
into the IMF. The main result, of course, is that in the large $p^+$
expansion the structure functions indeed have support only in the 
interval $x\in [0,1[$. The transformation (\ref{fboost}) was already 
obtained in ref. \cite{Ja80} for the $1+1$ dimensional bag model. 
{}From the observation that even in the realistic case the DIS is 
essentially $1+1$ dimensional, Jaffe then conjectured that the same 
relation would be true for $3+1$ dimensions as well. In contrast to 
the $1+1$ dimensional model of ref. \cite{Ja80}, however, a boost to 
the IMF must be implemented to satisfy the condition $x^+=0$. In the 
$1+1$ dimensional model this is a natural choice\footnote{In ref.
\cite{Be87} a comparison of the projection (\ref{fboost}) and
the method of Peierls--Yoccoz has been reported for the case of the
$1+1$--dimensional bag model. Only slight differences originating
from the sharp bag boundary were observed.}.

Finally it is important to note that the transformation (\ref{fboost}) 
receives additional (phenomenological) support from the na\"\i ve 
spectator model. In the rest frame one finds for large $x$ (actually 
larger than one) $f_{\rm RF}(x)\approx {\rm exp}(-\kappa N_Cx)$ 
\cite{Di96}. Hence the transformation (41) leads to structure functions
which drop like $f_{\rm IMF}(x)\to (1-x)^{\kappa N_C-1}$ as $x$ 
approaches unity. The constant of proportionality ($\kappa$) could 
eventually be determined numerically and compared to the spectator 
model result $(1-x)^{2N_s-1}$ where $N_s$ refers to the number of 
spectators. In this respect, in the large $N_C$ limit $N_s$ can be 
identified with the number of colors, $N_C$. Effectively, this 
comparison plays the role of a consistency check to the transformation
(\ref{fboost}).

\bigskip
\leftline{\large\bf 5. Numerical Effects}
\medskip

In the previous section we have verified Jaffe's conjecture
\cite{Ja80} that structure functions, which are calculated from
a localized $3+1$ dimensional quark configuration, get modified
according to eq (\ref{fboost}) when Lorentz covariance is restored
in the direction defined by the incident (virtual) photon by boosting 
to the IMF. Here we want to briefly
demonstrate with the help of an example that this not only provides
the proper support of the structure function but also effects the
predicted structure functions at small and moderate $x$. In figure
1 we therefore compare the valence quark approximation to
the unpolarized structure function $F_1(x)$ in deep--inelastic
electron nucleon scattering as it arises in the NJL chiral soliton
model \cite{We96}. The full structure function $F_1(x)$ also contains
the contribution of the backward moving intermediate quark,
$F_1^{(-)}(x)$, which is transformed according to eq (\ref{fboost})
as well.
\begin{figure}[t]
\vspace{1.0cm}
\centerline{
\epsfig{figure=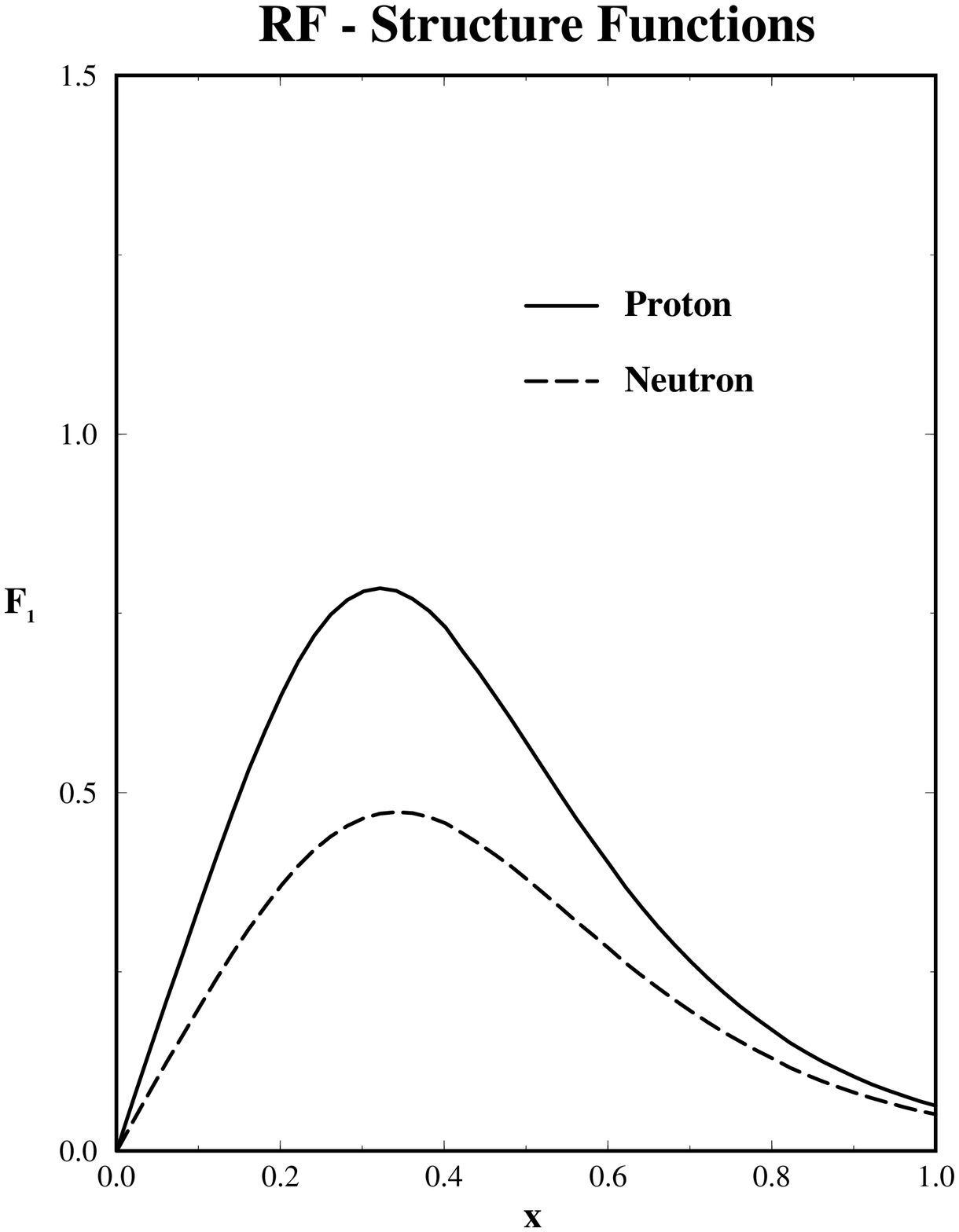,height=7.0cm,width=7.5cm}
\epsfig{figure=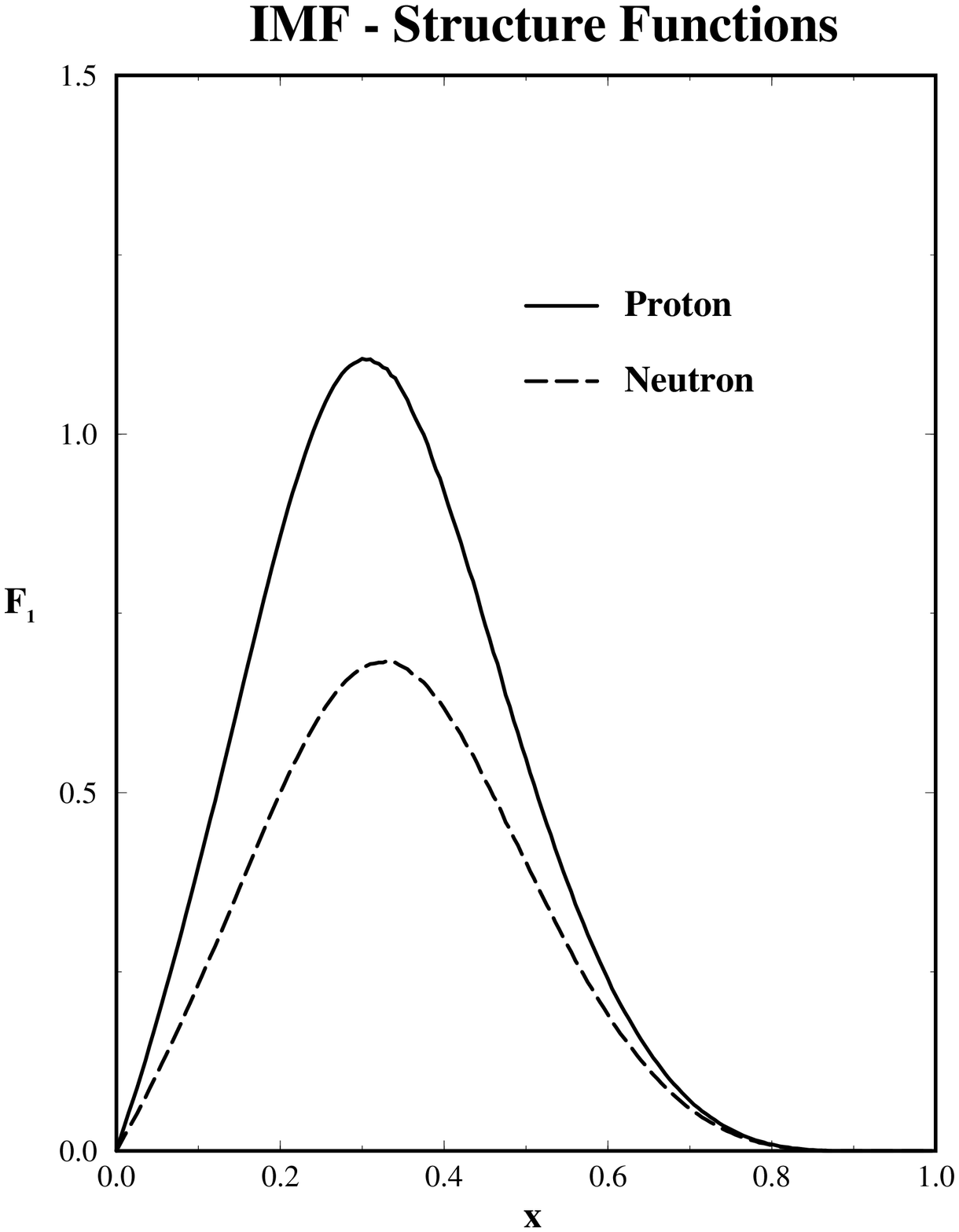,height=7.0cm,width=7.5cm}}
\vspace{0.5cm}
\caption{\label{fig_1}The unpolarized structure functions $F_1(x)$ for 
deep--inelastic electron nucleon scattering as functions of 
the Bjorken variable $x$. Left panel: rest frame calculation 
of ref. \protect\cite{We96}, right panel: projection according 
to eq \protect(\ref{fboost}).}
\end{figure}
Apparently the IMF structure functions drop to zero already at
$x\approx0.8$. On the other hand they are more pronounced in the
intermediate range $x\approx0.4$. This is, of course, a consequence
of the fact that the transformation (\ref{fboost}) conserves the area
under the curve.

\bigskip
\leftline{\large\bf 6. Conclusions}
\medskip

The calculation of nucleon structure functions in the Bjorken limit
singles out the hyperplane $\xi^+=0$. We have seen that upon transformation
to the IMF this condition can be satisfied even for models where
the nucleon emerges as a (static) localized object. For quark soliton
models this transformation can be performed by introducing a collective
coordinate which parameterizes the position of the soliton and
subsequently defining a boost in the space of the this coordinate.
Elevating this coordinate as well as its conjugate momentum to
operators in the framework of the semiclassical quantization not only
generates states of good momentum but  allows one to restore Lorentz
covariance, at least in the $\xi^\pm$ subspace. Fortunately this is
sufficient when calculating structure functions because those
components of the nucleon momentum which are orthogonal to this
subspace may be put to zero. When computing the structure functions in
a static soliton model with the quark spinors, which reside in the
background of the soliton, boosted to the IMF we have observed that the
common problem of improper support for the structure functions,
{\it i.e.} non--vanishing structure functions for $x>1$, is cured
along the line suggested by Jaffe \cite{Ja80} some time ago. The reason
is that the Lorentz contraction associated with the boost to the IMF
maps the infinite line exactly onto the interval $x\in [0,1[$.
Furthermore for the case of electron--nucleon scattering we have seen
that this Lorentz contraction effects the structure functions also
at small and moderate $x$.

\bigskip
\leftline{\large\bf Acknowledgements}
\medskip
One of us (LG) is grateful to G. R. Goldstein for helpful comments
and to K. A. Milton for encouragement and support.

\appendix
\stepcounter{chapter}
\bigskip
\leftline{\large\bf Appendix: Boosting to the Infinite Momentum Frame}
\bigskip

In this appendix we detail the two step process of restoring proper 
support to nucleon structure functions. We recall that the RF contribution 
of the forward moving intermediate quark to the unpolarized spin 
structure function, $F_{1}^{(+)}\left(x,q^2\right)$ is given in terms of 
rest--frame quark wave functions \cite{Ja75,Hu77,We96},
\be
F^{(+)}_1\left(x,q^2\right)\hspace{-0.1cm}&=&\hspace{-0.1cm}
-\lim_{q^-\to\infty}\
\frac{q^\rho}{4\pi}\int d\xi^+ d\xi^- d^2\xi_\bot 
\frac{\delta\left(\xi^2 \right)\epsilon(\xi^+ + \xi^-)}{2\pi}\
{\rm e}^{i\left(q^+\xi ^-+q^-\xi^+\right)}
\nonumber \\ && \hspace{-1cm} \times
\int \frac{dx_0^- d^2x_{0\bot}}{\sqrt 2}\ 
\langle N |\Psi^\dagger_+\left(\xi-x_0\right)\Gamma^\rho
\Psi_+\left(-x_0\right)|N\rangle\ ,
\label{f1p0}
\ee
where $|N\rangle$ is related to the zero momentum nucleon state 
\cite{Ja75,Hu77,Ba79}, 
\be
|\mbox{\boldmath $p$}=0\rangle=
\left[(2\pi)^3 2m\right]^{\frac{1}{2}}|N\rangle\ .
\ee
In eq (\ref{f1p0}) the integration over the collective coordinate $x_0$ 
corresponds to averaging over the position of the localized quark fields.
In the next step we perform the Lorentz transformation of the quark wave 
functions from the rest frame to the IMF specified by $\Lambda$ and 
eqs (\ref{rapidity}) and (\ref{dirtrans}), 
\be
\Psi_{x_0}(\xi)\quad {\stackrel{\scriptstyle{\rm IMF}}
{\textstyle\longrightarrow}}\quad
\Psi\left({\rm e}^\Omega(\xi^-_0-x^-_0),
\mbox{\boldmath $\xi$}_{0\bot}-\mbox{\boldmath $x$}_{0\bot},
\xi^+ - x^+_0=0\right) S\left(\Lambda_{\Omega}\right) \ .
\ee
Substituting this boosted quark wave function into eq (\ref{f1p0})
yields the Lorentz transformed expression of eq (\ref{f1p0}),
\be
F^{(+)}_1\left(x,q^2\right)
\hspace{-0.12cm}&=&\hspace{-0.12cm}-\hspace{-0.12cm}
\lim_{q^-\to\infty}\frac{q^\rho}{4\pi}
\int d\xi^+d\xi^-d^2\xi_\bot\ \frac{dx_0^-d^2x_{0\bot}}{\sqrt 2}
\frac{\delta(\xi^2)\epsilon(\xi^+ + \xi^-)}{2\pi}
{\rm e}^{i\left(q^+\xi^-+q^-\xi^+\right)}
\nonumber \\ && \hspace{1.0cm}\times
\Big\langle p^+, p^-=\frac{m}{\sqrt{2}p^+},
\mbox{\boldmath $p$}_\bot=0\Big| x_{0}\Big\rangle
\nonumber \\ && \hspace{1.0cm}\times
\langle N|\Bigg[\Psi\left({\rm e}^\Omega(\xi^- -x^-_0),
\mbox{\boldmath $\xi$}_\bot-\mbox{\boldmath $x$}_{0\bot},
\xi^+-x_0^+=0\right)S\left(\Lambda_{\Omega}\right)\Bigg]^\dagger 
\Gamma^\rho
\nonumber \\ && \hspace{4.0cm}\times
\Psi\left({\rm e}^\Omega(-x^-_0),
-\mbox{\boldmath $x$}_{0\bot},x^+_0=0\right)
S\left(\Lambda_{\Omega}\right)\Big|N\rangle
\nonumber \\ && \hspace{1.0cm}\times
\langle x_{0}|p^+, p^-=\frac{m}{\sqrt{2}p^+},
\mbox{\boldmath $p$}_\bot=0\Big\rangle\ ,
\ee
which yields eq (\ref{f1p1}). In addition we have identified the
nucleon wave function in terms of the coollective coordinate $x_0$.
It is most convenient to treat the
collective coordinate with the help of Dirac--$\delta$ functions
as it allows us to extract the dependence on the rapidity
${\rm e}^\Omega=\sqrt{2}p^+/m$ 
\be
\Psi\left(\frac{\sqrt{2}p^+}{m}(\xi^--x^-_0),
\mbox{\boldmath $\xi$}_\bot-\mbox{\boldmath $x$}_{0\bot},
\xi^+-x^+=0\right) &&
\nonumber \\&& \hspace{-5cm} 
=\frac{m}{\sqrt2\ p^+}
\int\ d\zeta^-\ d\zeta^2_\bot\
\delta\left(\xi^- - x_0^- - \frac{m}{\sqrt2\ p^+}\zeta^-\right)
\nonumber \\ &&  \hspace{-3.0cm} \times
\delta^2\left(\mbox{\boldmath $\xi$}_\bot-
\mbox{\boldmath $x$}_{0\bot}-
\mbox{\boldmath $\zeta$}_\bot\right)
S\left(\Lambda_{\Omega}\right)\Psi(\zeta)
\nonumber \\ && 
\nonumber \\ && \hspace{-5cm}
=\frac{m}{\sqrt2} \int\ d\zeta^-\ d\zeta^2_\bot
\frac{d\alpha}{2\pi}
{\rm exp}\left[i\alpha\left(p^+\xi^-+\Lambda-
\frac{m}{\sqrt2}\zeta^-\right)\right]
\nonumber \\ &&  \hspace{-3.0cm} \times
\delta^2\left(\mbox{\boldmath $\xi$}_\bot-
\mbox{\boldmath $x$}_{0\bot}-
\mbox{\boldmath $\zeta$}_\bot\right)
S\left(\Lambda_{\Omega}\right)\Psi(\zeta)\ .
\ee
Apparently the dependence on the nucleon momentum $p^+$ has 
disappeared from the argument of the quark wave--function $\Psi(\zeta)$.
This technique has repeatedly been used in section 4.

\end{document}